\documentclass{emulateapj}
\usepackage{apjfonts}
\usepackage{lscape}
\input{epsf}

\slugcomment{To be submitted to the Astrophysical Journal Letters}

\shorttitle{New ultra-cool subdwarfs from SDSS}
\shortauthors{L\'epine \& Scholz}

\begin{document}

\title{Twenty-three new ultra-cool subdwarfs from the Sloan Digital
  Sky Survey.\altaffilmark{1}}

\author{S\'ebastien L\'epine\altaffilmark{2}, \& Ralf-Dieter
  Scholz\altaffilmark{3}}

\altaffiltext{1}{The SDSS is managed by the Astrophysical Research
  Consortium for the Participating Institutions. The Participating
  Institutions are the American Museum of Natural History,
  Astrophysical Institute Potsdam, University of Basel, University of
  Cambridge, Case Western Reserve University, University of Chicago,
  Drexel University, Fermilab, the Institute for Advanced Study, the
  Japan Participation Group, Johns Hopkins University, the Joint
  Institute for Nuclear Astrophysics, the Kavli Institute for Particle
  Astrophysics and Cosmology, the Korean Scientist Group, the Chinese
  Academy of Sciences (LAMOST), Los Alamos National Laboratory, the
  Max-Planck-Institute for Astronomy (MPIA), the Max-Planck-Institute
  for Astrophysics (MPA), New Mexico State University, Ohio State
  University, University of Pittsburgh, University of Portsmouth,
  Princeton University, the United States Naval Observatory, and the
  University of Washington.}
\altaffiltext{2}{Department of Astrophysics, American Museum of
  Natural History, Central Park West at 79th Street, New York, NY
  10024, USA}
\altaffiltext{3}{Astrophysical Institute Potsdam, An der Sternwarte
  16, D-14481, Potsdam, Germany}

\begin{abstract}
A search of the Sloan Digital Sky Survey spectroscopic database has
turned up 23 new ultra-cool subdwarfs, low-mass metal-poor stars of
spectral subtype M 7.0 or later. Spectra from these red objects all
show very strong molecular bands of CaH but relatively weak bands of
TiO, indicative of a cool, metal-poor atmosphere. Five of the stars are
formally classified as M subdwarfs (sdM7.0-sdM8.5), 13 as more
metal-poor extreme subdwarfs (esdM7.0-esdM8.0), and five as extremely
metal-poor ultra subdwarfs (usdM7.0-usdM7.5). In the [$H_r$,$r-z$]
reduced proper motion diagram, these subdwarfs clearly populate the
locus of low-luminosity stars with halo kinematics. It is argued that
the objects are all very low-mass, metal-poor stars from the Galactic
halo (Population II). These new discoveries more than double the
census of spectroscopically confirmed ultra-cool subdwarfs. We show
that the stars stand out remarkably in the [g-r,r-i] color-color
diagram; a proposed color and proper motion selection scheme is expected
to be extremely efficient in identifying more of these old, very
low mass stars in the vicinity of the Sun.
\end{abstract}


\section{Introduction}

Ultra-cool dwarfs are very red objects with spectral subtype M7.0 or
later. The class includes very low mass stars, with masses just above
the hydrogen-burning limit, and extends into the substellar (brown dwarf)
regime. While ultra-cool dwarfs are expected to be common, their low
luminosities make them very challenging objects to find and study,
outside of the immediate vicinity of the Sun. The census of
spectroscopically confirmed ultra-cool dwarfs remains quite small
compared with the very large numbers of low-mass stars now formally
classified as M0-M6 dwarfs \citet{W08}. 

Ultra-cool {\em subdwarfs} are the metal-poor analogs of the
ultra-cool dwarfs. While representatives of the latter group have been
known for some time \citep{KHM91,KHS95}, it is only recently that
ultra-cool subdwarfs have been spectroscopically identified. In the
solar vicinity, ultra-cool subdwarfs are kinematically associated with
the Galactic halo \citep{L03a}, and represent the low-mass end of the
local Population II. As such, their local density is quite low, which
explains their scarcity in the solar vicinity.

Dwarfs and subdwarfs of spectral type M have optical spectra dominated
by molecular bands of metal hydrides (CaH, FeH) and metal oxides (TiO,
VO). Under the classification system recently proposed by
\citet{LRS07}, that expands the system of \citet{G97}, main
sequence M stars are segregated into four ``metallicity
classes.'' These are assigned on the basis of the relative strength of oxide
and hydride bands, which dominate the optical-red spectrum and whose
ratio is believed to be a function of log Z \citep{GR97}. Solar
metallicity objects are classified as dwarfs (M, or dM), while the
more metal-poor stars are classified as subdwarfs (sdM), extreme
subdwarfs (esdM), and ultra subdwarfs (usdM), in order of decreasing
metallicity. At the turn of the century, only two ultra-cool subdwarfs
were known: the sdM7.0 subdwarf LHS 377 \citep{G97}, and the esdM7.0
extreme subdwarf APMPM J0559-2903 \citep{S99}. Since then, a dozen
more objects have been discovered, spanning the spectral type range
sdM7.0-sdM9.5, esdM7.0-esdM8.5, and usdM7.0-usdM7.5
\citep{L03b,L04,S04b,G06,BK06,B07,LRS07}. 

In addition, metal-poor stars have also been found into the
L spectral type regime \citep{B03,B04}. These objects are clearly
substellar in nature, and believed to be Population II brown dwarfs. A
case in contention is the star LSR J1610-0040, initially classified as
an sdL \citep{L03c}, but which may be a peculiar sdM
\citep{CV06,RB06}. One thing is clear at this time: the census of
local ultracool subdwarfs is largely incomplete, and our knowledge
fragmentary. Much more extensive samples are required, which would be
critical in constraining the low end of the Galactic halo mass
function, and understanding the properties of cool, metal-poor
atmospheres.

This Letter reports the discovery of 23 new ultracool subdwarfs from
the Sloan Digital Sky Survey (SDSS) spectroscopic database. This more
than doubles the current census of spectroscopically confirmed
ultracool subdwarfs, and suggests a simple method for the
identification of more such stars. The search and identification is
described in \S2, where the spectra are also presented. Colors and
reduced proper motions are analyzed in \S3. Results are discussed in
\S4. 

\begin{deluxetable*}{lrrrrrrrrr}
\tabletypesize{\scriptsize}
\tablecolumns{10} 
\tablewidth{500pt} 
\tablecaption{New, spectroscopically confirmed ultra-cool subdwarfs} 
\tablehead{designation & $\mu$R.A.& $\mu$Decl. & g & r & i & z & J\tablenotemark{a} & spec.type \\
 & mas yr$^{-1}$&  mas yr$^{-1}$& mag & mag & mag & mag & mag & }
\startdata
SDSS J020533.75+123824.0 & -36.1$\pm$02.8&-271.0$\pm$06.3& 21.95$\pm$0.10& 19.76$\pm$0.02& 18.11$\pm$0.01& 17.30$\pm$0.01& 15.91$\pm$0.07&  sdM 7.5\\ 
SDSS J023557.61+010800.5 & -20.3$\pm$10.6&-202.5$\pm$01.8& 22.39$\pm$0.11& 20.39$\pm$0.03& 19.19$\pm$0.01& 18.55$\pm$0.03& \nodata       & esdM 7.5\\
SDSS J080301.12+354848.5 & +21.2$\pm$04.0&-489.1$\pm$09.3& 22.07$\pm$0.08& 20.19$\pm$0.02& 19.11$\pm$0.01& 18.33$\pm$0.02& \nodata       & esdM 7.5\\
SDSS J081210.47+372321.1 &-125.9$\pm$07.8&-182.2$\pm$09.7& 21.70$\pm$0.04& 19.71$\pm$0.01& 18.41$\pm$0.01& 17.75$\pm$0.01& 16.43$\pm$0.13& esdM 7.0\\
SDSS J101200.28+204611.6 &-304.1$\pm$04.6&-247.6$\pm$05.7& 21.90$\pm$0.06& 19.68$\pm$0.01& 18.26$\pm$0.00& 17.46$\pm$0.01& 16.21$\pm$0.07& esdM 7.5\\
SDSS J102757.77+340146.8 & -83.2$\pm$06.7&-523.8$\pm$10.5& 22.26$\pm$0.10& 20.34$\pm$0.02& 18.58$\pm$0.01& 17.67$\pm$0.01& 16.16$\pm$0.10&  sdM 8.0\\
SDSS J102839.48+593908.2 &+108.5$\pm$07.0&-323.1$\pm$06.5& 22.52$\pm$0.11& 20.39$\pm$0.03& 19.09$\pm$0.01& 18.29$\pm$0.03& \nodata       & usdM 7.5\\
SDSS J105047.16+242241.9 & +70.7$\pm$04.1&-455.5$\pm$03.4& 22.49$\pm$0.09& 20.31$\pm$0.02& 18.85$\pm$0.01& 18.09$\pm$0.02& \nodata       & esdM 8.0\\
SDSS J105449.01+322527.0 & +18.0$\pm$08.9&-125.5$\pm$01.8& 22.71$\pm$0.12& 20.54$\pm$0.03& 19.41$\pm$0.01& 18.76$\pm$0.03& \nodata       & usdM 7.0\\
SDSS J105717.29+462102.3 & -32.8$\pm$16.6&-528.1$\pm$14.1& 21.98$\pm$0.07& 19.99$\pm$0.02& 18.61$\pm$0.01& 17.89$\pm$0.02& \nodata       & esdM 7.0\\
SDSS J110651.29+044814.9 & -89.9$\pm$06.4&-282.0$\pm$03.5& 22.74$\pm$0.15& 20.56$\pm$0.03& 18.80$\pm$0.01& 17.81$\pm$0.02& 16.29$\pm$0.10&  sdM 8.5\\
SDSS J114452.67-031420.4 & +26.2$\pm$05.7&-254.6$\pm$08.4& 22.84$\pm$0.18& 20.66$\pm$0.04& 19.51$\pm$0.02& 18.68$\pm$0.04& \nodata       & esdM 7.0\\
SDSS J123257.20+410723.5 & +43.1$\pm$11.2&-182.3$\pm$05.4& 23.36$\pm$0.21& 20.95$\pm$0.04& 19.84$\pm$0.02& 19.14$\pm$0.05& \nodata       & esdM 7.0\\
SDSS J125508.35+294939.1 &-188.2$\pm$03.9& -67.5$\pm$02.4& 22.51$\pm$0.11& 20.38$\pm$0.03& 19.26$\pm$0.02& 18.68$\pm$0.04& \nodata       & usdM 7.0\\
SDSS J125938.97+493404.2 &-155.2$\pm$17.7& -65.8$\pm$05.8& 22.50$\pm$0.09& 20.56$\pm$0.02& 19.27$\pm$0.01& 18.50$\pm$0.02& \nodata       & esdM 7.0\\
SDSS J130852.73+391142.8 &-304.4$\pm$02.9&-258.9$\pm$17.4& 22.98$\pm$0.11& 20.94$\pm$0.03& 19.32$\pm$0.01& 18.46$\pm$0.02& \nodata       &  sdM 7.0\\
SDSS J135128.49+550656.9 &-245.5$\pm$02.3&-166.4$\pm$01.5& 21.17$\pm$0.03& 18.99$\pm$0.01& 17.68$\pm$0.00& 16.92$\pm$0.01& 15.68$\pm$0.06& esdM 7.0\\
SDSS J141053.86+621742.6 &-155.1$\pm$09.2& +50.5$\pm$07.2& 22.52$\pm$0.10& 20.47$\pm$0.03& 19.19$\pm$0.01& 18.50$\pm$0.03& 16.92$\pm$0.16& esdM 8.0\\
SDSS J143645.25+310914.7 &-155.4$\pm$07.1& -99.3$\pm$03.9& 22.25$\pm$0.08& 20.29$\pm$0.02& 19.04$\pm$0.01& 18.31$\pm$0.03& \nodata       & esdM 7.0\\
SDSS J153037.83+495205.7 &-214.4$\pm$05.5& -12.3$\pm$05.1& 22.54$\pm$0.09& 20.56$\pm$0.02& 18.97$\pm$0.01& 18.14$\pm$0.02& 16.86$\pm$0.18&  sdM 7.0\\
SDSS J155707.09+283531.5 & -58.0$\pm$10.4&-217.9$\pm$32.5& 22.76$\pm$0.11& 20.51$\pm$0.02& 19.28$\pm$0.01& 18.61$\pm$0.03& \nodata       & usdM 7.0\\
SDSS J160347.10+164145.3 & -72.9$\pm$04.6&-105.4$\pm$05.5& 22.95$\pm$0.13& 20.74$\pm$0.03& 19.52$\pm$0.02& 18.88$\pm$0.05& \nodata       & usdM 7.5\\
SDSS J164451.75+310352.8 & -31.0$\pm$13.9&-105.8$\pm$17.1& 23.05$\pm$0.16& 20.85$\pm$0.04& 19.44$\pm$0.02& 18.66$\pm$0.03& \nodata       & esdM 8.0  
 \enddata
\tablenotetext{a}{Infrared J magnitudes from the 2MASS All-Sky Catalog
  of Point Sources\citep{C03}}
\end{deluxetable*}

\section{Spectroscopic identification}

A search for cool subdwarfs was performed on the spectroscopic
database of the Sloan Digital Sky Survey \citep{G98,Y00}. A subsample of
44,600 SDSS spectra from the sixth data release \citep{DR6} was
retrieved, including all spectra from point-like sources (STAR and
STAR LATE flags) with colors $g-r>1.0$. Each spectrum was
automatically matched against a grid of composite spectra of M dwarfs
and subdwarfs assembled by L\'epine (2008, in preparation). This
grid, built on the 4-class metallicity system of \citet{LRS07},
includes reference spectra for M stars of all four metallicity
subclasses (dM, sdM, esdM, usdM), covering all known spectral subtypes
with a 0.5-subtype resolution. The new grid expands and complements
the M dwarf classification templates assembled by
\citet{Bo07}.

Spectral classes and subtypes are assigned on the basis of a least-squares
fit over the full 6100-7900\AA spectral range. This improves on the
standard classification scheme for M subdwarfs, which is based on the
absolute and relative strengths of the three line indices CaH2, CaH3,
and TiO5 \citep{RHG95,G97,L03a,LRS07}. The main problem with those
molecular line indices is that they are defined over relatively narrow
spectral ranges, making them unreliable in low signal-to-noise
spectra. A fit over a much broader spectral region made it possible to
classify SDSS spectra from very faint sources, with a signal-to-noise
ratio per resolution element as low as S/N$\sim5$. Since the
classification grid is built from a combination of those same SDSS
spectra, several iterations were made on which the grid was
modified, and vetted against the spectral-index classification system.

\begin{figure*}
\epsscale{1.05}
\plotone{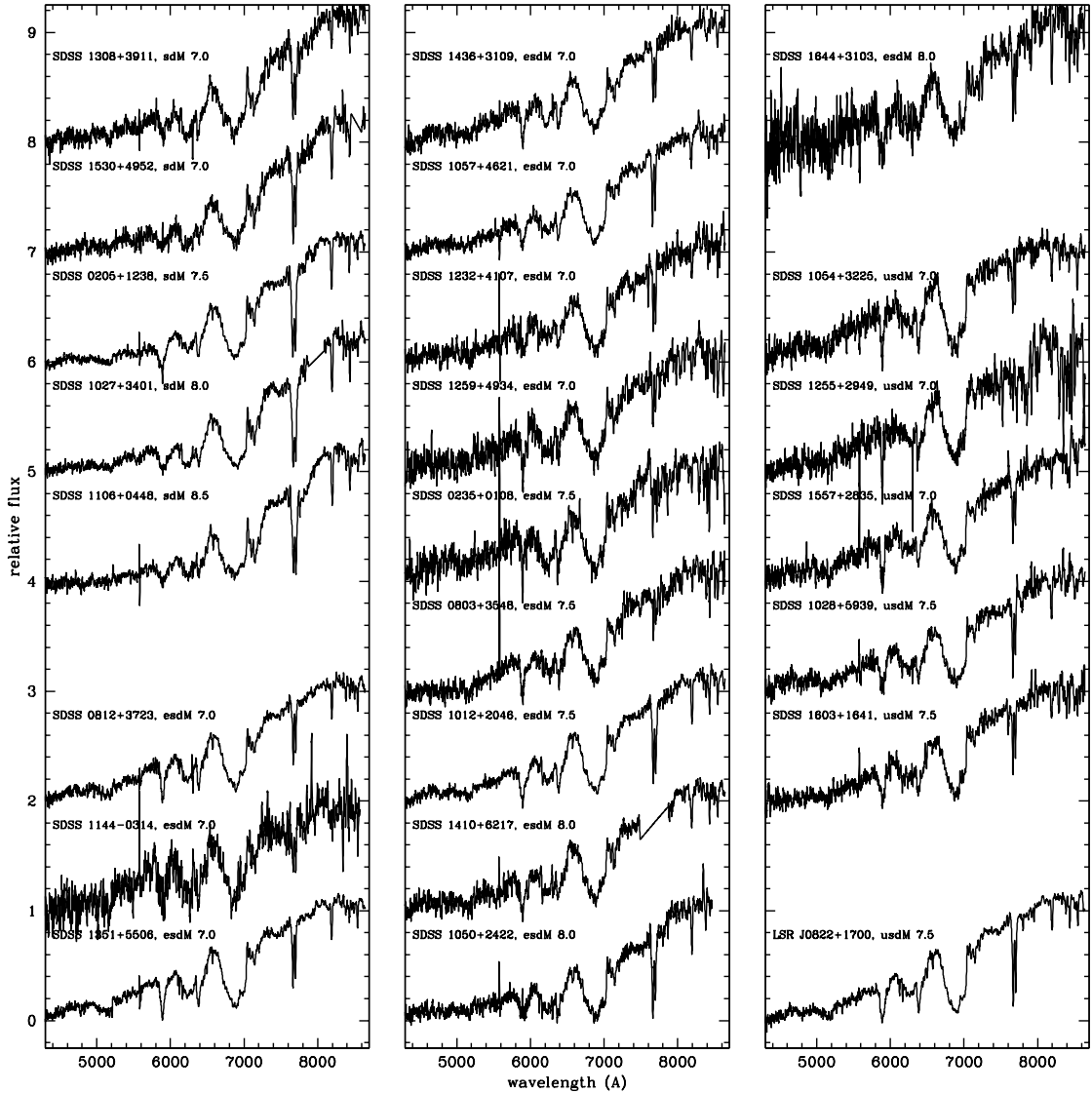}
\caption{SDSS medium-resolution spectra of all ultra-cool subdwarfs
  identified in the SDSS database, all new objects except for the
  usdM7.5 classification standard LSR J0822+1700 ({\it bottom right}). The
  most prominent feature in all those very red spectra is the deep CaH
  molecular absorption band just blueward of 7,000\AA. The strength of
  the TiO bands redward of 7,000\AA is used to assign metallicity
  classes, as the ratio of TiO to CaH is assumed to be proportional
  to log Z.}
\end{figure*}

Our automated analysis of the 44,600 red stellar spectra in the end
yielded the formal identification of 24 M subdwarfs of spectral
subtype 7.0 or later. One of the spectra was found to be of the nearby
high proper motion star LSR J0822+1700, recently selected as a
subdwarf classification standard (usdM7.5) by \citet{LRS07}. All the
other stars are identified here as ultra-cool subdwarfs for the first
time. The final tally includes five subdwarfs (sdM7.5-sdM8.5), 13 extreme
subdwarfs (esdM7.0-esdM8.0) and five ultra subdwarfs
(usdM7.0-usdM7.5). Spectra of all 23 new ultra-cool subdwarfs are
presented in Figure 1, along with the spectrum of LSR J0822+1700. 

Basic astrometry and photometry is presented in Table 1, along with
SDSS {\it griz} photometry \citep{G98,Fetal96}, and J magnitude from
the 2MASS All-Sky Catalog of Point Sources \citep{C03}. All except one
(SDSS J1351+5506) are fainter than any of the previously known
ultra-cool M subdwarfs; 15 of them are even too faint to show up in
2MASS. Proper motions were found in the SDSS catalog for only a few
stars. We instead determined the proper motions for all the stars from
their photographic plate counterparts, as found in the SuperCOSMOS sky
survey \citep{H01} and the APM-North catalogue \citep{MM00}. Proper
motions were calculated from a linear regression of the early-epoch
plate positions and later SDSS coordinates. Results are shown in Table
1.

\section{Colors and reduced proper motion}

Figure 2 shows the location of our ultra-cool subdwarfs in the
[$g-r$,$r-i$] diagram ({\it left panel}). For comparison, a
subsample of 50,000 random SDSS point sources is also displayed
({\it dots}). The ultra-cool subdwarfs are found to stand out significantly
from the locus of the field stars, confirming the trend observed
for earlier M subdwarfs by \citet{W04}. Stars are also segregated by
metallicity subclass, with the sdM having larger $r-i$ colors and
usdM smaller $r-i$ values. The trend suggests that at lower
metallicities, cool stars tend to fall more in line with the
extrapolated color-color sequence of the more massive stars ({\it
  dashed line}), consistent with a spectral energy distribution more
similar to that of a blackbody.

\begin{figure*}[t]
\epsscale{1.05}
\plottwo{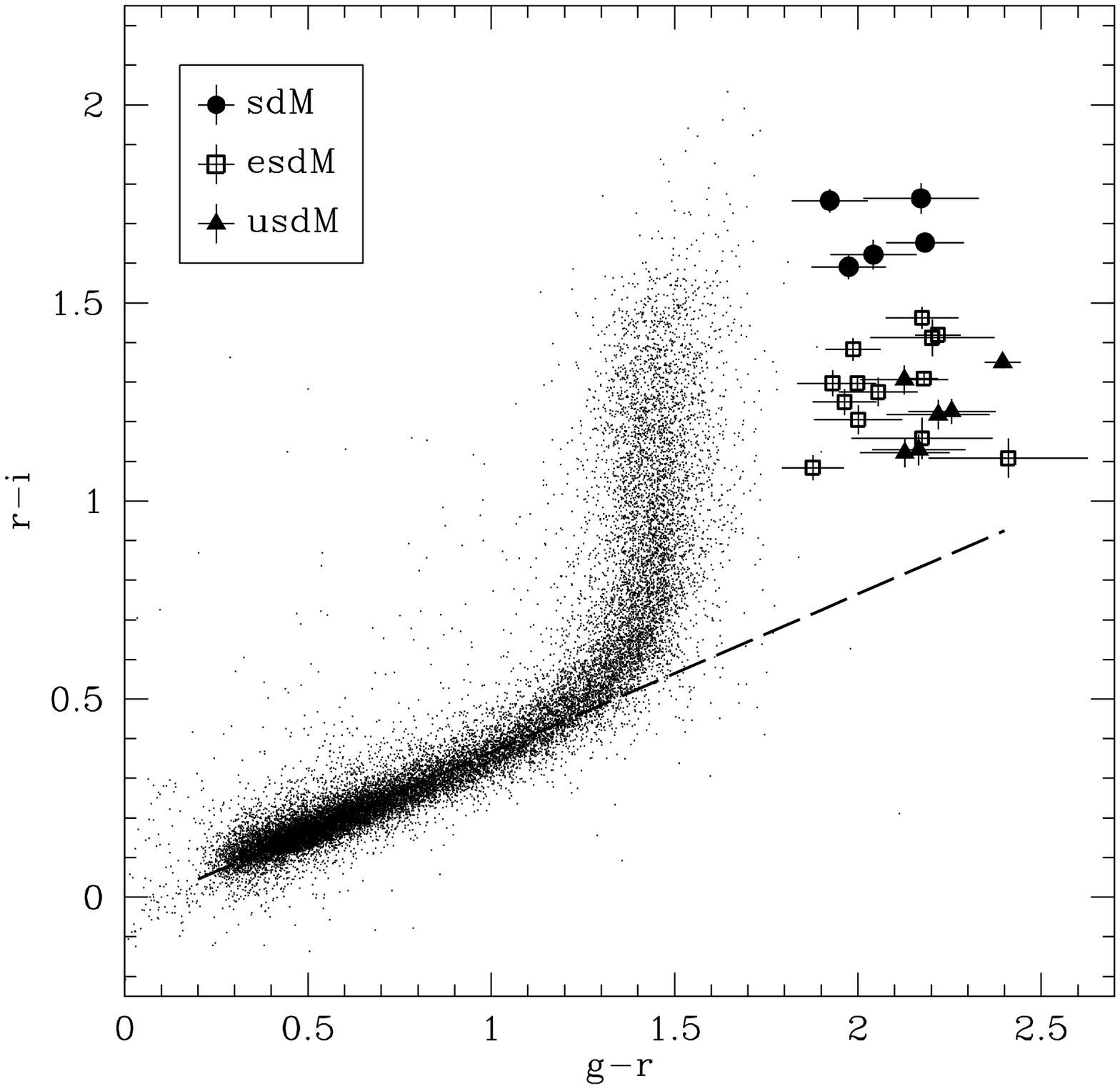}{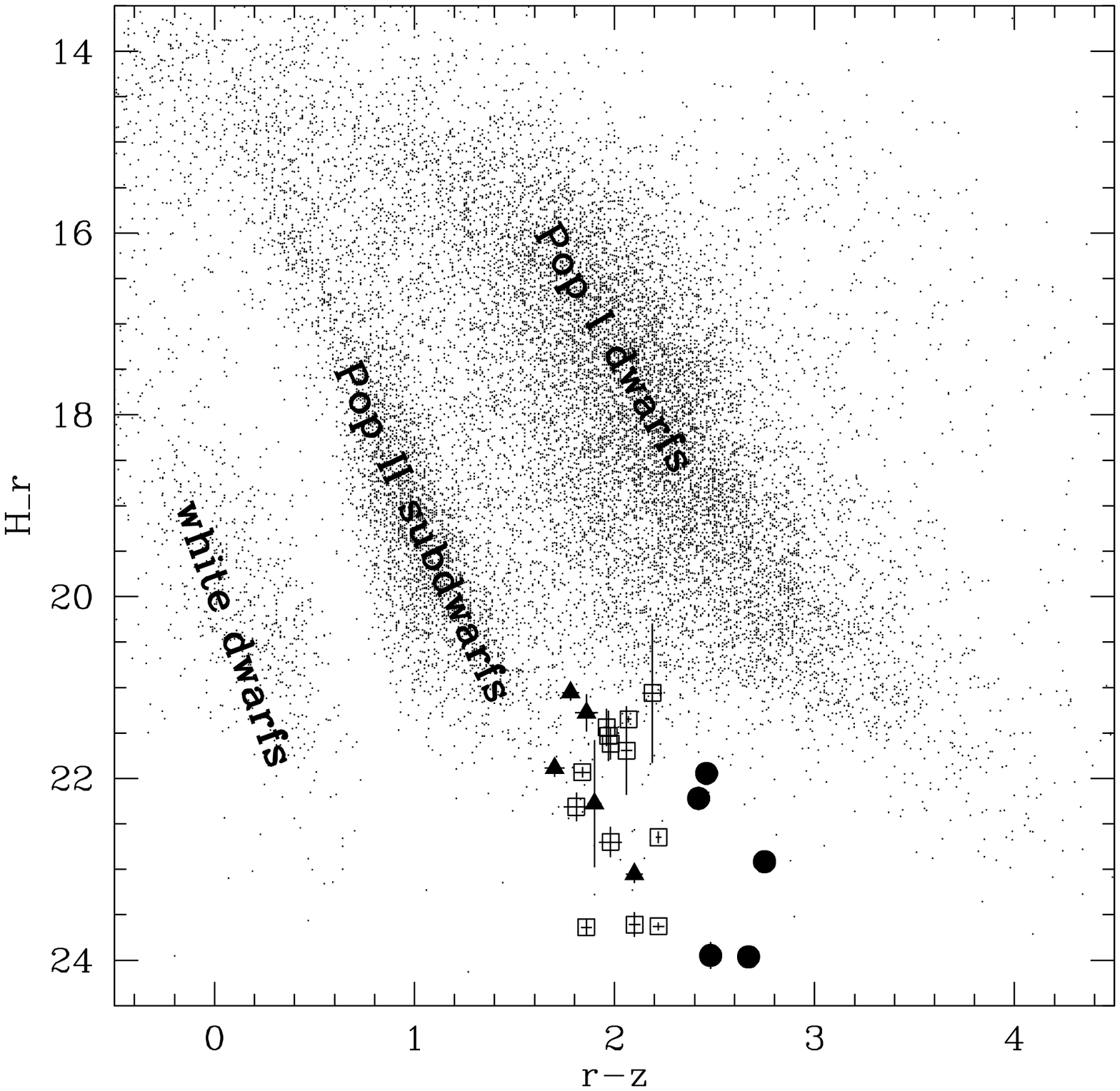}
\caption{
{\it Left panel:} Color-color magnitude diagram [$g-r$,$r-i$] of SDSS
ultra-cool subdwarfs ({\it symbols}). The subdwarfs are clear outliers to the
standard locus of field stars, here denoted with a random sample
of 50,000 SDSS point sources ({\it dots}). The ultra-cool subdwarfs are also
clearly segregated by spectral class, which suggests a stellar locus
strongly dependent on metallicity, from the strong ``elbow-shaped''
locus for stars of solar abundances to an almost ``straight-line''
locus at very low metallicities. {\it Right panel}: Reduced proper
motion diagram of the SDSS ultra-cool subdwarfs ({\it symbols}). The stars
populate the low-luminosity end of the Population II main sequence,
which unambiguously identifies them as low-mass subdwarfs. A
distribution of high proper motion stars from the LSPM-North catalog
\citep{LS05} with SDSS counterparts is shown for comparison ({\it dots});
the distinct loci of disk dwarfs, halo subdwarfs, and white dwarfs (in
layers from upper right to lower left, respectively) are clearly
defined.}
\end{figure*}

Stars for which proper motions are available all have $\mu>200$ mas
yr$^{-1}$, which suggests that they are either high-velocity objects
and/or have very low luminosities. The right panel in Figure 2 shows the
[$H_r$,$r-z$] reduced proper motion diagram for all the SDSS
ultra-cool subdwarfs with recorded $\mu>200 mas yr^{-1}$. The objects
are all located along the extension of the Population II sequence
\citep{Y03,LRS07}, which runs between the Population I sequence ({\it
  upper right}) and the white dwarf sequence ({\it lower left}). This
unambiguously demonstrates that the stars are low-mass objects at the
bottom of the main sequence and kinematically associated with the
Galactic halo population.

The very distinctive distribution of colors and reduced proper motions
suggests an efficient method to identify ultra-cool subdwarfs. They
should be found among stellar sources with $g-r>1.85$ and
$g-i>3.1$ (see Fig.2). Additional proper motion or reduced proper
motion constraints would optimally select for those low-luminosity
objects, with minimal background contamination. This would work best
at high Galactic latitudes, where interstellar reddening is minimal.

\section{Discussion and conclusions}

Our successful search for ultra-cool subdwarfs demonstrates that these
stars can be identified in deep optical surveys like SDSS. The stars
stand out in the (g-r,r-i) color-color diagram, their locus clearly
distinct from that of solar-metallicity objects. This makes them easy
to identify from point sources in fields imaged in the SDSS {\it
  ugriz} color system. It is expected that ultra-cool subdwarfs will
be easy to identify in future, deep imaging surveys, provided they use
similar {\it gri} bandpasses. The fact that ultra-cool subdwarfs are
segregated by metallicity in color-color space also makes them highly
valuable tracers of Galactic halo structure and history.

In color-color space, the ultracool subdwarfs populate a region
typically occupied by extra-galactic objects such as QSOs
\citep{Retal02}. Indeed, all 23 new ultra-cool subdwarfs presented in
this Letter were QSO targets in the SDSS spectroscopic
survey. Interestingly, this is not the first time that ultra-cool
stars are discovered as a by-product of QSO surveys; ultra-cool dwarfs
were also identified in early surveys of high-redshift quasars
\citep{K97}. But just like subdwarfs are a contaminant for QSO
surveys, QSOs would also be a common contaminant of subdwarf surveys.
Absolute proper motion cuts would be necessary to limit the
contamination from extragalactic sources. It is fortunate that subdwarfs
are kinematically associated with the local Galactic halo population
and tend to be high-velocity stars in the vicinity of the Sun. A
combination of color cuts and minimal proper motion requirement would
be the most efficient technique to select ultra-cool subdwarfs.

In the meantime, the spectra collected here will be used to further
refine the classification templates, and assist in the identification
and classification of ultra-cool subdwarfs in upcoming SDSS data
releases. The subdwarf classification routine is now being integrated
to the SEGUE spectral reduction pipeline, and ultra-cool subdwarfs are
now expected to be routinely identified in the SDSS survey.

\acknowledgments

{\bf Acknowledgments}

This research was supported by NSF grant AST-0607757 at the American
Museum of Natural History. Funding for the SDSS and SDSS-II has been
provided by the Alfred P. Sloan Foundation, the Participating
Institutions, the National Science Foundation, the U.S. Department of
Energy, the National Aeronautics and Space Administration, the
Japanese Monbukagakusho, the Max Planck Society, and the Higher
Education Funding Council for England. The SDSS Web Site is
http://www.sdss.org/.


\end{document}